# Progressive observation of Covid-19 vaccination effects on skin-cellular structures by use of Intelligent Laser Speckle Classification (ILSC)

Ahmet Orun
De Montfort University, Faculty of Computing, Engineering and Media, Leicester UK, LE1 9BH
Email: aorun@dmu.ac.uk.  Phone: +44(0)116 3664408

Fatih Kurugollu
University of Derby, School of Computing and Engineering
Derby, UK.  Email:  f.kurugollu@derby.ac.uk

## Abstract

We have made a progressive observation of Covid-19 Astra Zeneca Vaccination effect on Skin cellular network and properties by use of well established *Intelligent Laser Speckle Classification* (*ILSC*) image based technique and managed to distinguish between three different subjects groups via their laser speckle skin image samplings such as early-vaccinated, late-vaccinated and non-vaccinated individuals. The results have proven that the ILSC technique in association with the optimised Bayesian network is capable of classifying skin changes of vaccinated and non-vaccinated individuals and also of detecting progressive development made on skin cellular properties for a month period.

## 1 Introduction

Covid-19 infection or its vaccine's effect on the skin structure and components have been studied and proven by several authors (Sun et al., 2021; Magro et al., 2021; Bogdanov et al.,2021) as Covid-19 is a systemic disease whose signs are also reflected by the skin. Hence an effective and speedy diagnosis and detection of such systemic diseases could only be done by a non-destructive instant skin analysis using optical imaging methods (Orun et al.,2014; Orun, 2014) rather than microscopic inspection with biopsies that cannot be non-destructive and in real-time. In the previous studies, Magro et al. (2021) made a comprehensive analysis of the Covid-19 effects on skin by using light microscopy for skin biopsies but their method does target neither an instant disease diagnosis nor an observation of its prognosis. In another work, Bogdanov et al. (2021) focused on the adverse effects of the Covid-19 vaccine on the skin. Their research is based on basic visual inspection or biopsy techniques. Almost all other global studies on Covid-19 vaccines effect (Lospinoso et al., 2021) rely mainly on visual inspections and biopsies which require highly equipped clinical environments. Within this research, we propose an alternative non-destructive optical (instant) observation technique of Covid-19 vaccine effects called *ILSC* which could also be potentially used for the diagnosis and prognosis of the disease.

The laser speckle technique has several advantages over the other traditional optical methods (Millet et al., 2011) for skin analysis such as, due to its specifically selected wavelength smaller than skin cells, the laser light can interact with the cellular network to detect its changes as a textural phenomenon rather than focusing on a single cell (Orun et al., 2014).  This ability provides a power of high sensitivity and spatial resolution to observe a network of very small features like cells or even sub-cellular features (e.g. cell's nuclei, etc.).  The advantage of lasers over an ordinary broad spectrum (non-coherent) light sources are well known characteristics (Ginouves, 2003) , this is due to the micro level interaction of lasers with skin features and its deep penetration into biological tissues. The proposed system also consists of basic level instrumentation (e.g. camera, laser source, etc.) and is highly competitive against high-cost and non-real-time methods like electron microscopy, optical tomography, ultrasound, Raman microscopy, etc.

With the addition of  AI capability to laser speckle imaging technique which is called "Intelligent laser speckle classification" *(ILSC)*, the method becomes more efficient for a textural features classification. *ILSC*  is an integrated technique [1][2][3][4][5] (Orun et al., 2014a, 2014b, 2016, 2017)  which comprises different disciplinary methods like laser light-object interaction (laser speckle physics), image texture analysis and Artificial Intelligence (AI) methods (e.g. Bayesian Networks) which provides high sensitivity and classification power to detect subtle differences even at the subcellular level. The method detects laser speckle textural differences, hence it does not focus on an individual skin cell, but rather on the whole cellular network to detect any property or structural changes in all cells simultaneously as all cells are naturally identical to each other. To increase the physical detection sensitivity most suitable laser wavelength has to be selected. In our case red laser ($\lambda$=650 nm) is used as the optimum one for human skin and blood cells analysis (Butkus, 2003). The most important advantage of the method is its non-invasive real-time continuous observation by a camera-laser source base simple system. Figure 1 depicts the image acquisition setup configuration which is used to acquire Laser speckle images of the cellular network. The laser wavelength $\lambda$= 0.65 µm is shorter than individual cell size (~30µm) or sub-cellular components size (e.g. nucleus) So that the laser can interact and backscatter by conveying the characteristic pattern of cellular components to the speckle image domain. The surface normal makes 2º angles with a laser beam and camera viewing axis.

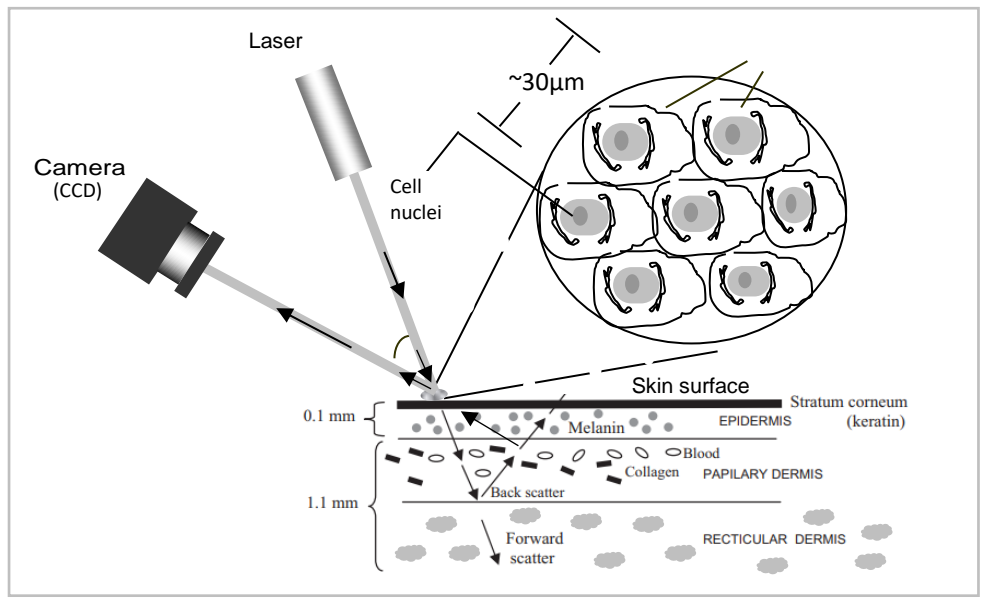

Figure.1 Image acquisition setup configuration with the basic system components

Like some of the most important diseases (e.g. diabetes, heart disease, kidney disorder, etc.) The covid-19 disease is also a systemic disease whose effect may be reflected by skin (Mohamed et al., 2021) and that is a supportive argument for this study that the Covid-19 vaccination progressive effect which is normally invisible to the human eye and causing cellular structural changes, an increase of antibodies in the blood, etc. may be observable by *ILSC* technique. According to the AstraZeneca[TM] report (AstraZeneca, 2021), a single dose of vaccination leads to a four-fold increase in antibodies in 95% of participants in one month after injection. Hence the tests used in this study covers one month period after the first injection.

The main contributions of this research to the earlier studies would be as follows:

- Within this research, we have proven that the Covid-19 vaccine effect and possibly Covid-19 infection can be observable at high frequency of data collection, with high reliability. Its reliability can be further increased by an advanced version of this proposed technique called "Multi-Classifier System" (MCS) by the addition of more laser sources with different wavelengths (e.g. red, IR, etc.) (Orun and Smith, 2017; Wozniak et al., 2014; Ho et al., 1994).
- Speedy image acquisition and processing provide a real-time skin analysis which enabling a continuous skin observation by successive data collection.
- Non-invasive characteristics by the use of low-level (< 1mW) laser light energy.
- Texture based laser speckle image classification provides very high spatial resolution with "scalability" for the analysis of sub-cellular features (depending on specifically selected laser wavelength)
- The Laser speckle effect is a physical phenomenon which is very sensitive against the subtle changes, whose potential is exploited within this work.

- Even though the classification capability of the system is proven by Bayesian networks here, such AI components can easily be replaced by other methods (e.g. deep learning – Convolutional NN, etc.).
- Skin penetration of the laser light would also be depending on its selected wavelength. This provides a filtering capability of unwanted skin layers' obstruction for the ideal image data acquisition (Bashkatov, 2005).
- The low-cost and highly portable characteristics of the system is promising for its broad usage in the health community, transport environments, leisure centres, schools, etc.

## 2 materials and Methods

### 2.1 Laser speckle imaging camera system

For the purpose of image data acquisition, a high resolution (3840 x 2880 pixels) commercial CCD image camera is used where each CCD pixel corresponds to 2.8 µm$^2$ area on the image domain. The camera is utilized after a moderate modification by attaching a low level laser source (1mW) with approximately 23° of angle to the skin surface normal (Fig. 1). The laser source is collimated red laser diode at 650 nm wavelength [6] whose power is much less than maximum permissible exposure which is 2000 W m$^{-2}$ for human skin (for 400–700 nm wavelength range) for a long term exposure such as between 5–10 hours. The laser source is used in our tests only for short-term illuminations (approx. 3–5 s) over approximately 10 mm diameter area on the upper-hand skin of each participant to generate a speckle effect on the skin surface and sub-skin layers.

### 2.2 Laser speckle phenomenon

According to the basic principle of "laser light-surface interaction"; when a rough surface is illuminated by a coherent light like laser then the light itself scatters from the surface exhibiting a particular intensity distribution as it looks covering the surface with a granular structure. The very fundamental formula of laser speckle image includes its pixel intensity statistics (1), where the standard deviation of spatial intensity variations $\sigma_s$ is equal to the mean intensity ⟨I⟩ for a fully developed (ideal) speckle pattern. This would be stated by the basic formula;

$$K = \sigma_s / \langle I \rangle \qquad (1)$$

where K is the speckle contrast and its value takes place between 0 and 1. If the speckle pattern is ideal then K=1. But if the speckle pattern becomes not ideal such as blurred by a diffuser or surface motion, the value of K will be shifted towards zero. The laser speckle effect image formation and its statistical analysis at the skin observation domain by use of a digital camera (e.g. CCD ideally or CMOS at lower cost) is a multi-parametric task. The laser speckle image's (LSI) statistical property heavily depends on its system geometry like laser illumination angle or camera orientation. In the image formation domain a basic light amplitude at point A (e.g. at a single pixel of CCD camera matrix) may be formulized as in Formula 2 and 3 (Orun and Alkis, 2003);

$$I_j(A) = |I_j| \, e^{i\varphi_j} \qquad (2)$$

The complex amplitude $I_{com}$ in the image domain at point A would be written as ;

$$I_{com}(A) = \frac{1}{\sqrt{N}} \sum_{j=1}^{N} |I_j| e^{i\varphi_j} \qquad (3)$$

where;
$I_j$ : Basic light amplitude of a surface element j
A : Point on image domain (single pixel of CCD matrix)
$\varphi_j$ : Random phases of light at j$^{th}$ surface element.
i : Imaginary part
N : Total surface elements

## 2.3 Texture analysis of LSI

Image texture analysis (Phillips, 1995) utilizing specific texture measures, is an essential part of the Intelligent laser speckle classification (ILSC) system that was tested and used successfully in the last nearly 20 years with high accuracy (Orun and Smith, 2017; Orun and Alkis, 2003). The formulations are described in the Formulas 4-8. The nine types of texture measures used based on the Formulas 4-8, five ones with 3x3 pixel image scanning kernel, and four texture measures with 5x5 pixel kernel based on all of those formulas. The texture analysis is applied on 30x30 pixel image segment of each laser speckle image where the laser speckle effect appears to be as maximum (Figure 2). Sampling process of the laser speckle image areas has a great importance where the uniform textural characteristics of image segment should be justified. To ptimize this task, each 30x30 pixel sampling area should have sufficient number of texture primitives and also textural uniformity all over the area (Orun et al., 2014).

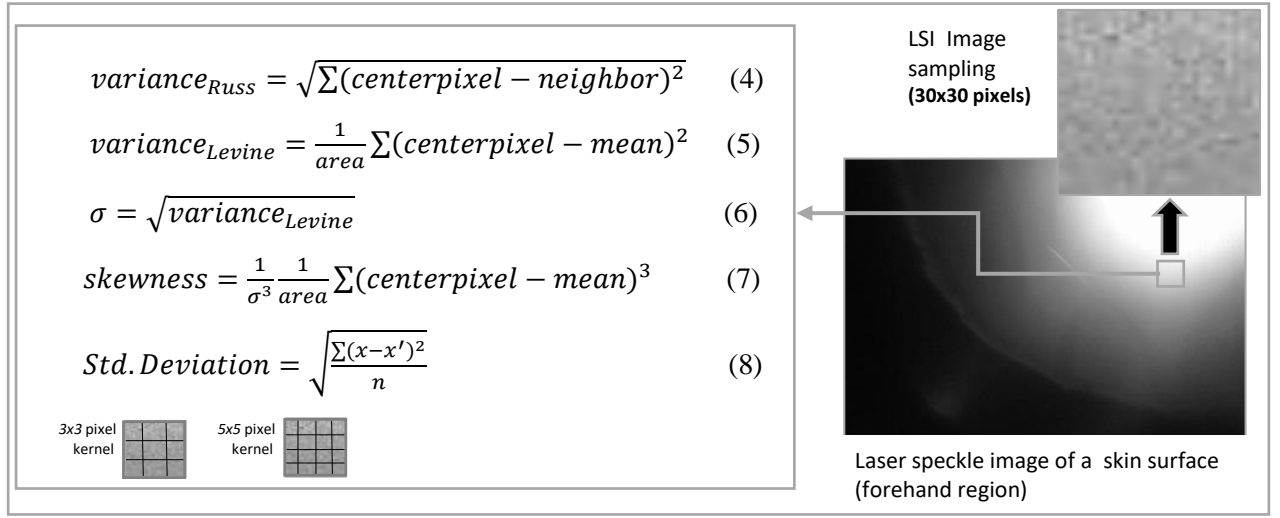

$$variance_{Russ} = \sqrt{\sum(centerpixel - neighbor)^2} \quad (4)$$

$$variance_{Levine} = \frac{1}{area}\sum(centerpixel - mean)^2 \quad (5)$$

$$\sigma = \sqrt{variance_{Levine}} \quad (6)$$

$$skewness = \frac{1}{\sigma^3}\frac{1}{area}\sum(centerpixel - mean)^3 \quad (7)$$

$$Std.Deviation = \sqrt{\frac{\sum(x-x')^2}{n}} \quad (8)$$

Figure 2. Pixel based arithmetical operators (on the left) for Image texture are applied on each laser speckle skin image sampling. Each 30x30 pixel image sampling is selected on the global laser illuminated image where the physical laser speckle effects particularly exist as shown in the square region of interest (on the right)

## 2.4 Bayesian Network as AI method

Bayesian Network is well established Artificial Intelligent method whose details and statistical description are already given in several publications (Drugan and Wiering, 2010; Jensen; 1998; Orun and Seker, 2012) As a brief, Bayesian algorithm examines the information of two related variables from a data set and decides if two variables are dependent. It also investigates how close the relationship is between those variables. This information is called conditional mutual information of two variables Xi, Xj which may be denoted as:

$$I(X_i, X_j|C) = \sum_{x_i,x_j,c} P(x_i, x_j, c) \log \frac{P(x_i, x_j|c)}{P(x_i|c)P(x_i|c)} \quad (9)$$

In Equation 9, C is a set of nodes and c is a vector (one instantiation of variables in C). If $I(X_i, X_j /C)$ is smaller than a certain threshold t, then we can say $X_i$ and $X_j$ are conditionally independent.

# 3 Results and discussion

## 3.1 Classification test

The experimental stage includes Laser speckle image (LSI) data collection from three different subjects groups: early vaccinated, late vaccinated and no-vaccinated individuals for a month period. Camera images are collected from the forehand region at the specific time of the day (midnight). Each image sampling size was 30x30 pixels targeting a specific laser speckle effect region and a single image pixel corresponds to 2.8 microns. Image data set includes 60 cases in total as collected from three subject groups (20 cases for each group) for a month period.

The subject groups are labelled as follows: **S**: early vaccinated, **A**: late vaccinated and **E or EM**: no-vaccination. Laser speckle image samples taken are quantized by textural analysis (Phillips, 1995) and then build data sets are processed by Bayesian network for the classification after a training session (by 50/50 ratio of data). In the second type of tests, the link discoveries between the data attributes (texture measures of LSI) are found for progressive change detection. The results show that the parametrically optimized Bayesian classifier called PowerPredictor$^{TM}$ (e.g. by attribute links threshold, discretisation method, single-multi net options, etc.) is capable of discriminating between vaccinated and non-vaccinated individuals (80%-90% classification accuracy) and display the equality between vaccinated individuals (65% classification accuracy).

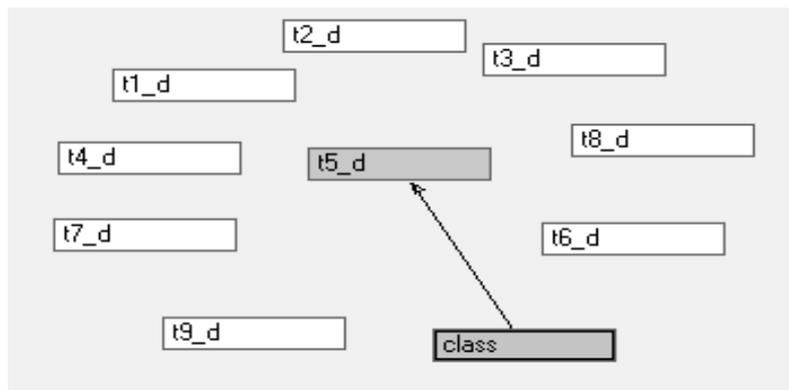

Figure 3. Classification results between the early vaccinated group (S) and the non-vaccinated group (EM) shown in the Bayesian network graph.

The classification has been achieved between the Group (S) and Group (EM) which are early vaccinated and non-vaccinated respectively with 80% classification accuracy results (sensitivity=100%, specificity=60%) (Automatically built network by PowerPredictor$^{TM}$ ) are presented in Figure 3 with the attribute nodes and connection links. In the network the attributes for Laser image texture measures are symbolized by $t_i$ where classification is achieved over the $5^{th}$ texture measure. The maximum classification accuracy has been achieved between the Group (A) and Group (EM) which are early vaccinated and non-vaccinated respectively with 90% classification accuracy (Sensitivity = 100%, Specificity = 80%) results as shown in Figure 4. To prove the identical effects of the vaccinations on cellular network structures, similar tests have been carried on with two groups of late vaccinated (group (A) and early vaccinated (group (S) with the results of 65% classification accuracy (Sensitivity = 80%,Specificity = 50%). This meant that the system (with the same network options) could not discriminate between those groups whose automatically constructed Bayesian network is shown in Figure 5.

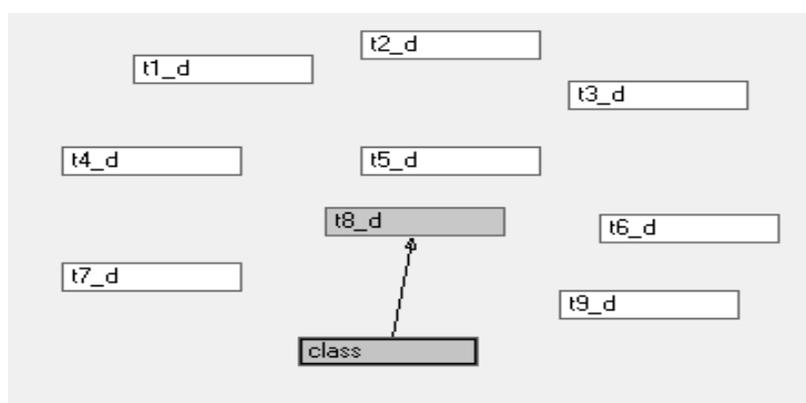

Figure 4. Classification results between the late vaccinated group (A) and the non-vaccinated group (EM) with 90% classification accuracy are obtained via automatically built network by PowerPredictor$^{TM}$. In the network the attributes for Laser image texture measures are symbolized by $t_i$ where classification is achieved over the $8^{th}$ texture measure (Skewness: 5x5 pixel kernel)

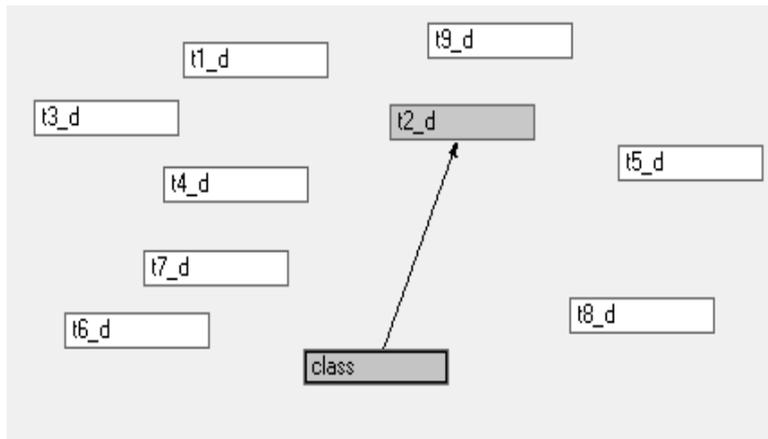

Figure 5. Classification results between the late vaccinated group (A) and the early vaccinated group (S) with 65% classification accuracy are obtained via automatically built network by PowerPredictor[TM]. It is shown that system can not discriminate between those vaccinated groups as expected.

## 3.2 Cellular Progressive Changes (CPC) observation test

In the progressive observation test, an artificial progression attribute (e.g. series of numbers between 1-60) is added to each data set to see whether any network link to any other vaccinated individuals' attribute will be established automatically by a specific Bayesian tool PowerConstructor[TM.] To prove that there is a progress of changes on a cellular network or cellular properties. This would be any change in cellular network structure in shape, an increase of antibodies in the blood (micro-veins) or any other cellular/subcellular content. In the results, the system establishes a link between the progression attribute and vaccinated groups (S, A) but not displays any link between the progression attribute and the non-vaccinated group as shown in Figure 6.

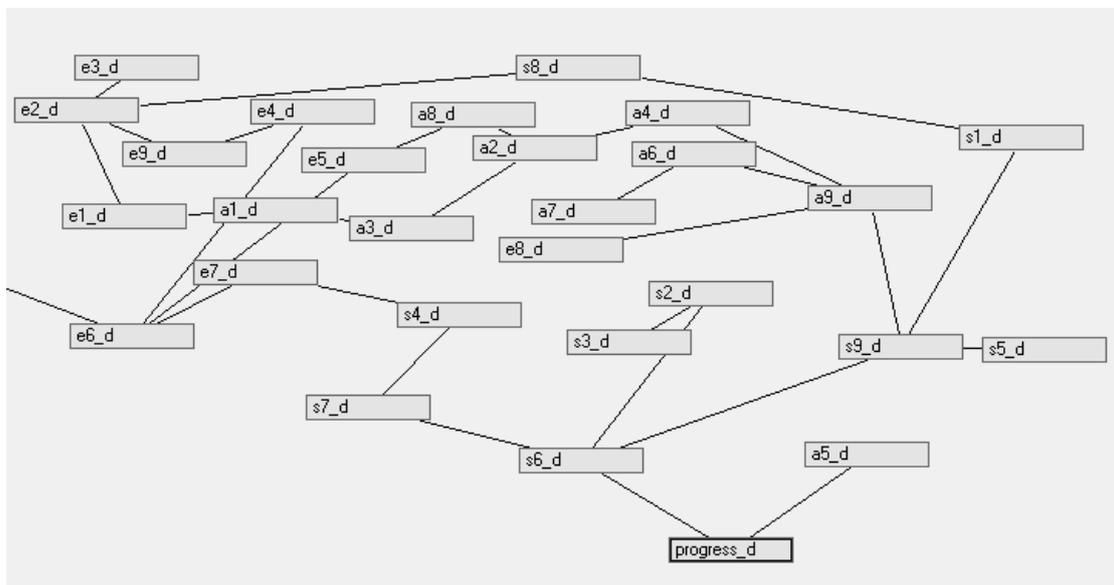

Figure 6. The Bayesian inference system (PowerConstructor[TM]) analyse the whole data set and finds a link between the "cellular progressive change" attribute and the vaccinated groups (**S**:early-vaccinated and **A**:late-vaccinated)

The Partial data set shown in Table 1 includes texture $t_1 – t_9$ calculations referring to pixel based statistics formulated in Figure 2 with two arithmetic operator (kernel) groups as are 3x3 and 5x5 pixels. Table 2 also includes an artificially built "progressive change" attribute in the last column to investigate if there is any link between any of the attributes and progression measures.

Table I. Partial data set including artificially built "progressive change" attribute shown in the last column with the regular increment (whole class column includes two classes)

| t1 | t2 | t3 | t4 | t5 | t6 | t7 | t8 | t9 | class | progress |
|---|---|---|---|---|---|---|---|---|---|---|
| 565 | 684 | 26.15 | 3.51 | 97 | 96 | 9.8 | 0.16 | 17.59 | E | 1 |
| 568 | 612 | 24.74 | 3.89 | 66 | 25 | 5 | 0.52 | 16.39 | E | 2 |
| 571 | 643 | 25.36 | 3.71 | 81 | 58 | 7.62 | 0.63 | 17.01 | E | 3 |
| 571 | 629 | 25.08 | 3.75 | 82 | 49 | 7 | 0.48 | 16.83 | E | 4 |
| 571 | 643 | 25.36 | 3.68 | 99 | 63 | 7.94 | 0.52 | 17.11 | E | 5 |
| 571 | 640 | 25.3 | 3.75 | 79 | 49 | 7 | 0.38 | 16.88 | E | 6 |
| 563 | 598 | 24.45 | 3.84 | 80 | 31 | 5.57 | 0.93 | 16.36 | E | 7 |
| 568 | 646 | 25.42 | 3.72 | 85 | 54 | 7.35 | 0.24 | 16.93 | E | 8 |
| 571 | 659 | 25.67 | 3.57 | 109 | 88 | 9.38 | 0.68 | 17.52 | E | 9 |
| 571 | 649 | 25.48 | 3.69 | 89 | 60 | 7.75 | 0.83 | 16.94 | E | 10 |
| 571 | 613 | 24.76 | 3.86 | 66 | 32 | 5.66 | 0.07 | 16.53 | E | 11 |
| 571 | 637 | 25.24 | 3.73 | 83 | 54 | 7.35 | 0.36 | 16.95 | E | 12 |
| 568 | 618 | 24.86 | 3.79 | 88 | 44 | 6.63 | 0.98 | 16.75 | E | 13 |
| 571 | 634 | 25.18 | 3.75 | 80 | 48 | 6.93 | 0.48 | 16.83 | E | 14 |

# Conclusion

The proposed potential use of Intelligent laser Speckle Classification and Bayesian Inference techniques are the unique ones among the other conventional methods like immunoglobin tests, as ILSC optimally unifies different disciplinary fields like light physics, texture analysis and AI technique to outperform the other conventional methods in terms of speed and non-invasiveness. The most common methods used to observe Covid-19 vaccines effects rely on the immunoglobin test which checks the amount of certain antibodies called immunoglobins generated progressively in the body. The antibodies are proteins that immune cells produce to fight against viruses. The proposed ILSC technique may also be utilized to observe the prognosis period of Covid-19 disease with the possible healing effect of particular substances (e.g. Dexamethasone (NHS England, 2021), Remdesivir (Dyer, 2020), etc.) at high frequency of tests. It can also be potentially used to select the most effective vaccine type for the individual (in the concept of personalised medicine (Vogenberg, 2010) by a frequent observation of the vaccine effect on the body tissue cells. The proposed system is getting more inevitable by its unique characteristics like its instant use even at the real-time speed as well as its flexible re-design capacity of instrumentation such as specifically selection of its laser wavelength suitable with the size of observation features (e.g. sub-cellular components), power of laser source to adjust its light penetration into the skin layers, camera resolution and another AI method used (e.g. Convolutional neural networks, etc.). The system reliability would be further increased by next advanced version of the proposed technique called "Multi-Classifier System" (MCS) by the addition of more laser sources with different wavelengths (e.g. red, IR, etc.). Further improvement of the system capability (e.g. blood cloth development risk assessment after vaccination) would depend on such data collection procedure from the specific patient group who suffer from blood cloth development issue, as technical configuration of proposed system would be capable enough to overcome such task with some minor system modifications like laser wavelength selection sensitive to subtle changes of blood content, circulation velocity, etc.